\documentclass[showpacs,preprintnumbers,amsmath,nofootinbib,aps]{revtex4}

\usepackage{graphicx,epsf, epsfig, amssymb, multirow}% Include figure files
\usepackage{bm}
\usepackage{amsmath}

\begin{document}
%\large
                                                                                                                           
\title{String resonances at the Large Hadron Collider}

\author{Arunava Roy}
\email{arunav@olemiss.edu}
\affiliation{Department of Physics and Astronomy, University of Mississippi,
University, MS 38677-1848, USA}
                                                                                                                           
\author{Marco Cavagli\`a}
\email{cavaglia@olemiss.edu}
\affiliation{Department of Physics and Astronomy, University of Mississippi,
University, MS 38677-1848, USA}

\date{\today}

\begin{abstract}
The Large Hadron Collider promises to discover new physics beyond the Standard Model. An exciting possibility is
the formation of string resonances at the TeV scale. In this article, we show how string resonances may be
detected at the LHC in the $pp\rightarrow\gamma+jet$ channel. Our study is based on event shape variables,
missing energy and momentum, maximum transverse momentum of photons and dijet invariant mass. These observables
provide interesting signatures which enable us to discriminate string events from the Standard Model background.
\end{abstract}
\maketitle 

\section{Introduction\label{intro}}
Our knowledge of high energy physics is limited to energies approximately less than one TeV. A possible candidate
for new physics above the TeV scale is supersymmetry (SUSY) \cite{Martin:1997ns,Bartl:1996dr}. SUSY  provides a
solution for the Higgs mass problem, a candidate for cold dark matter, and unification of low energy gauge
couplings by introducing superpartners to Standard Model (SM) fields (see Ref.\ \cite{Martin:1997ns} and
references therein). Alternatives to SUSY are extra-dimensional models, such as large extra
dimensions \cite{ArkaniHamed:1998rs}, warped braneworlds \cite{Randall:1999ee} and universal extra dimensions
\cite{Appelquist:2000nn}.  In these models, gravity becomes strong at the TeV scale. The most astounding consequences
of TeV-scale gravity would be the production of mini black holes (BHs) \cite{Argyres:1998qn} and  real/virtual
gravitons \cite{Giudice:1998ck} in particle colliders and cosmic ray showers. 

Both SUSY and extra dimensions are essential ingredients of string theory
\cite{Polchinski:1998rq,Zwiebach:2004tj}. The string scale is defined as \cite{Zwiebach:2004tj} 
\begin{equation*}
l_s=\bar{h}c\sqrt{\alpha^{\prime}},
\label{eq1}
\end{equation*}
where $\alpha^{\prime}$ is the slope parameter with units of inverse energy squared. The strength of string interactions
is controlled by the string coupling $g_s$. The Planck scale $M_{PL}$ is related to the string scale
$M_s$ by
\begin{equation*}
M_s=g_s M_{PL}.
\label{eq2}
\end{equation*}
Since string effects are expected to appear just before quantum gravity effects set in \cite{Dienes:1996du}, the
string coupling is generally assumed to be of order one. In this scenario, the string scale is close to the
Planck  scale. However, in the presence of large extra dimensions gravity becomes strong at the TeV scale. In
this case the Planck scale and the string scale are both $\sim$ 1 TeV; string resonances would be observed at the
Large Hadron Collider (LHC) before the onset of non-perturbative quantum gravity effects.  Detection of string
events at the LHC \cite{Burikham:2004su, Anchordoqui:2008ac, Anchordoqui:2008di} through corrections to SM
amplitudes would be the most direct evidence of this scenario. 
 
The aim of this article is to present a detailed analysis of string resonances at the LHC in the
$pp\rightarrow\gamma+jet$ channel \cite{Anchordoqui:2008ac}. (For a discussion of different channels see, e.g.,
Ref.~\cite{Burikham:2004su}.)  The main result of our investigation is that string resonances may be
distinguishable from the SM background.

\section{String Amplitude\label{sr}}
The relevant process for $pp\rightarrow\gamma+jet$ events is gluon-gluon scattering: $gg\rightarrow g\gamma$. The
string amplitude for  this process is \cite{Anchordoqui:2008ac}
\begin{equation}
|M(gg\rightarrow g\gamma)|^2=g_s^4 Q^2 C(N)\left\{\left[\frac{s\mu(s,t,u)}{u}+\frac{s\mu(s,u,t)}{t}\right]^2+(s\longleftrightarrow
t)+(s\longleftrightarrow u)\right\},
\label{string_amp}
\end{equation}
where $s$, $t$ and $u$ are the Mandelstam variables and
\begin{equation}
\mu(s,t,u)=\Gamma(1-u)\left(\frac{\Gamma(1-s)}{\Gamma(1+t)}-\frac{\Gamma(1-t)}{\Gamma(1+s)}\right).
\end{equation}
Here $N$=3 is the number of $D$ branes needed to generate the eight gluons of the SM,
$C(N)=\frac{2(N^2-4)}{N(N^2-1)}$ is a constant parameter, and $Q^2=\frac{1}{6}\kappa^2 \cos^2\theta_W\sim 2.55
\times 10^{-3}$, where $\kappa^2$=0.02 and $\theta_W$ are the mixing parameter and the Weinberg angle, respectively.
The values of the parameters are chosen as in Ref.~\cite{Anchordoqui:2008ac}.

The string amplitude possesses poles at $n$=$s/M_s^2$, where $n$ is an integer.
For odd values of $n$ the amplitude is
\begin{equation}
|M(gg\rightarrow g\gamma)|^2=g_s^4 Q^2
C(N)\frac{4}{(n!)^2}\frac{s^4+u^4+t^4}{M_s^4[s-nM_s^2]}\left\{\frac{\Gamma(t/M_s^2+n)}{\Gamma(t/M_s^2+1)}\right\}^2.
\label{string_amp1}
\end{equation}
For even values of $n$ the behavior of the amplitude is obtained from Eq.~(\ref{string_amp1}) with the
substitution $s\rightarrow t$ and $n\rightarrow m=t/M_s^2$ in the square bracket term. Following
Ref.~\cite{Anchordoqui:2008ac}, the singularities of the amplitude are smeared with a fixed width $\Gamma=0.1$ for
all $n>1$ and as
\begin{equation}
|M(gg\rightarrow g\gamma)|^2 \sim \frac{g^4 Q^2
C(N)}{M_s^4}\left\{\frac{M_s^8}{(s-M_s^2)^2+(\Gamma^{J=0} M_s^2)^2}
+\frac{t^4+u^4}{(s-M_s^2)^2+(\Gamma^{J=1} M_s^2)^2}\right\}
\label{eq4}
\end{equation}
for $n$=1. Equation~(\ref{eq4}) includes a correction for spin dependent widths: $\Gamma^{J=0}=0.75 \alpha_s M_s$
and $\Gamma^{J=1}=0.45 \alpha_s M_s$, where $\alpha_s=g_s^2/4\pi$ is the strong coupling constant. The presence of
the poles indicates the formation of string resonances.  The total cross section for the $pp \rightarrow
\gamma+jet$ event is obtained by integrating the parton cross section over the CTEQ6D parton distribution
functions of the protons $f(x,Q)$ \cite{Pumplin:2002vw}
\begin{equation}
\sigma_{pp \rightarrow string \rightarrow \gamma+jet}=\int_0^1 dx_1 \int_0^1 dx_2 \int_{t} dt~
f_1(x_1,Q) f_2(x_2,Q) \frac{d\sigma}{dt},
\end{equation}
where $Q$ is the four-momentum transfer squared and
\begin{equation}
\frac{d\sigma}{dt}=\frac{|M(gg\rightarrow g\gamma)|^2}{16 \pi s^2}.
\end{equation}
The choice of CTEQ6D parton distribution functions allows direct comparison of
our results to those of
Ref.~\cite{Anchordoqui:2008ac}.
\begin{figure*}[ht]
\centerline{\null\hfill
    \includegraphics*[width=0.50\textwidth]{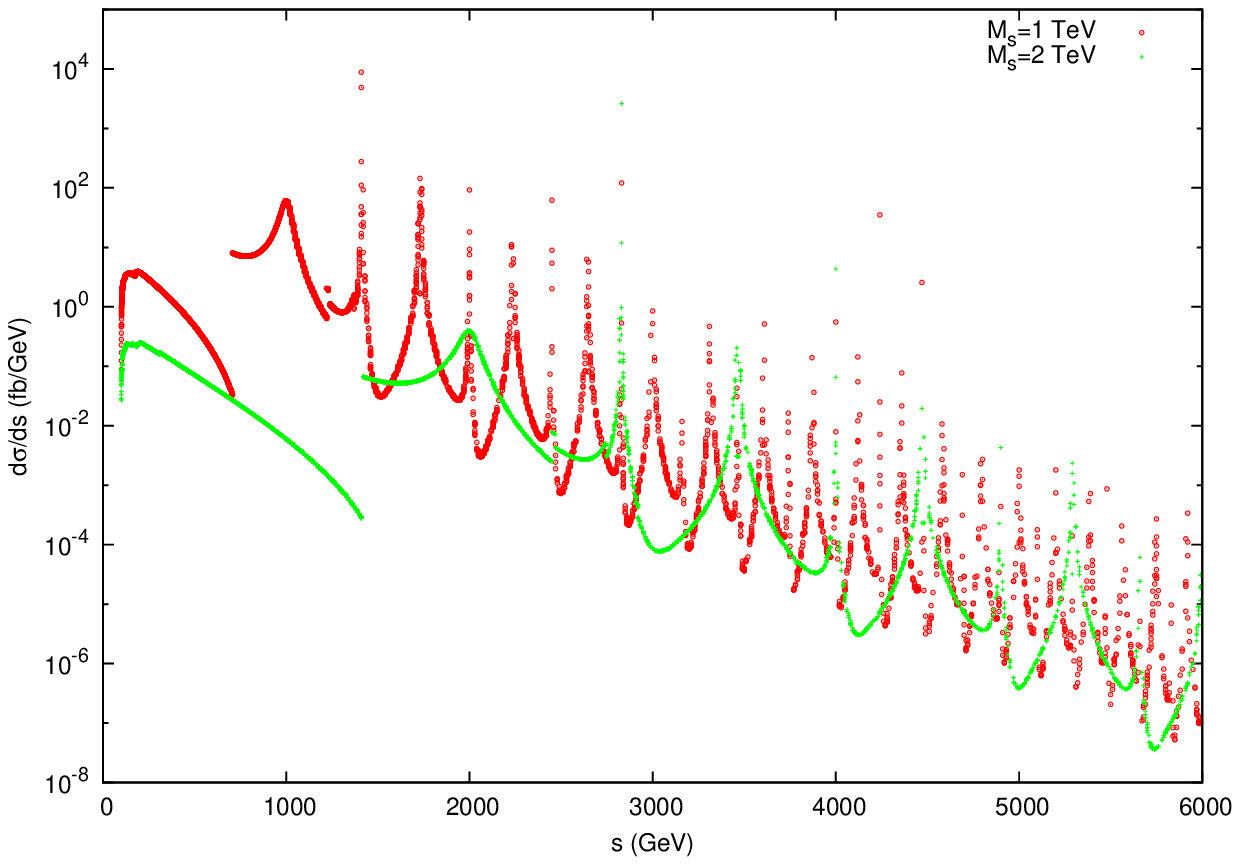}
    \null\hfill
    \includegraphics*[width=0.50\textwidth]{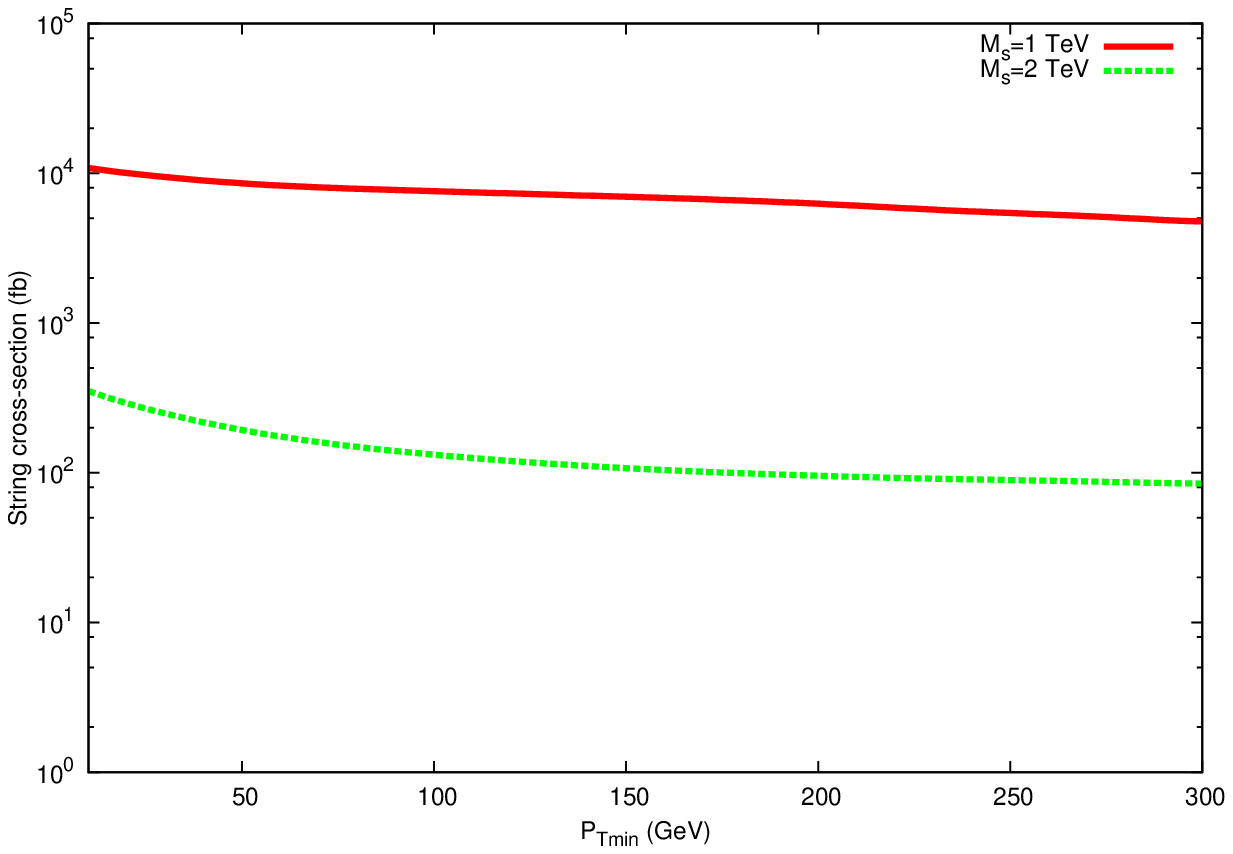}
    \null\hfill
    }
\caption{Left Panel: Differential cross section of string events for $M_s$= 1 TeV (red dots) and $M_s$= 2 TeV
(green crosses) with $P_{Tmin}$=50 GeV. String resonances are clearly seen when $s=nM_s^2$. Right Panel:
String cross section for $M_s$= 1 TeV (red dots) and $M_s$= 2 TeV (green crosses) with $P_{Tmin}$=50 GeV. The
cross section for $M_s$= 1 TeV is $\sim$ 44 times larger than the cross section for $M_s$= 2 TeV.}
\label{fig0}
\end{figure*}
The left panel of  Fig.~\ref{fig0} shows the differential cross section of the
$pp\rightarrow \gamma+jet$ process with total center-of-mass energy $E_{CM}$
\begin{equation}
\frac{d\sigma}{ds}=\int_{x_2} \int_{t} dx_2~dt \frac{2 \sqrt{s}}{x_2 E_{CM}^2} f_1(x_1,Q)
f_2(x_2,Q) \frac{d\sigma}{dt}.
\end{equation}
The differential cross section shows resonances at $s=n M_s^2$. The right panel of Fig.~\ref{fig0} shows the
total cross section as a function of the minimum transverse momenta of the two outgoing particles of the
$2\times2$ scattering, $P_{Tmin}$.  The string cross section for $M_s$=1 TeV (solid red line) and the cross
section for $M_s$=2 TeV (dashed green line) are $\sim$ 5$\times 10^4$ and $10^3$ times less than the SM cross
section, respectively. Our sample run of $10^7$ events produced $\sim$ 9300 (220) string events for $M_s$=1~(2)
TeV, with an integrated LHC luminosity of $100~fb^{-1}$.
\section{Analysis\label{analy}}
String resonances at the LHC are simulated with a Fortran Monte Carlo code interfaced with PYTHIA
\cite{Sjostrand:2006za}. Event-shape variables are a powerful discriminator of string events from the SM
background. Their effectiveness is further increased by an analysis of events with high-$P_T$ photons. String
events are characterized by high values of visible energy and missing transverse momentum as the photon and the
jet originate directly from the $2\times2$ interaction. Isolated photons provide a further means to extract string
signals. Being directly produced from the string resonance, isolated photons from string interactions are harder
than SM photons.

We fix $P_{Tmin}$=50 GeV for both string and SM events which results into a  signal-to-background ratio of $\sim$
73. This choice is justified as follows. For low values of $P_{Tmin}$ the string cross section is highly
suppressed $w.r.t.$ the SM cross section, for example $\frac{\sigma_{string}}{\sigma_{SM}}\sim10^{-5}$ for
$P_{Tmin}$~=~10 GeV. Therefore, discrimination of string events from the SM background is difficult for events
with low $P_{Tmin}$. At higher values of $P_{Tmin}$ both the SM background and the signal are substantially
reduced. For example, at 300 GeV they are reduced by a factor of $\sim$ 98\% and $\sim$ 42\% $w.r.t.$ values the
at $P_{Tmin}$~=~50 GeV, respectively. Thus the optimal signal-to-background ratio is obtained for $P_{Tmin}\lesssim$
100 GeV.
\begin{figure*}[ht]
\centerline{
    \null\hfill
    \includegraphics*[width=0.5\textwidth]{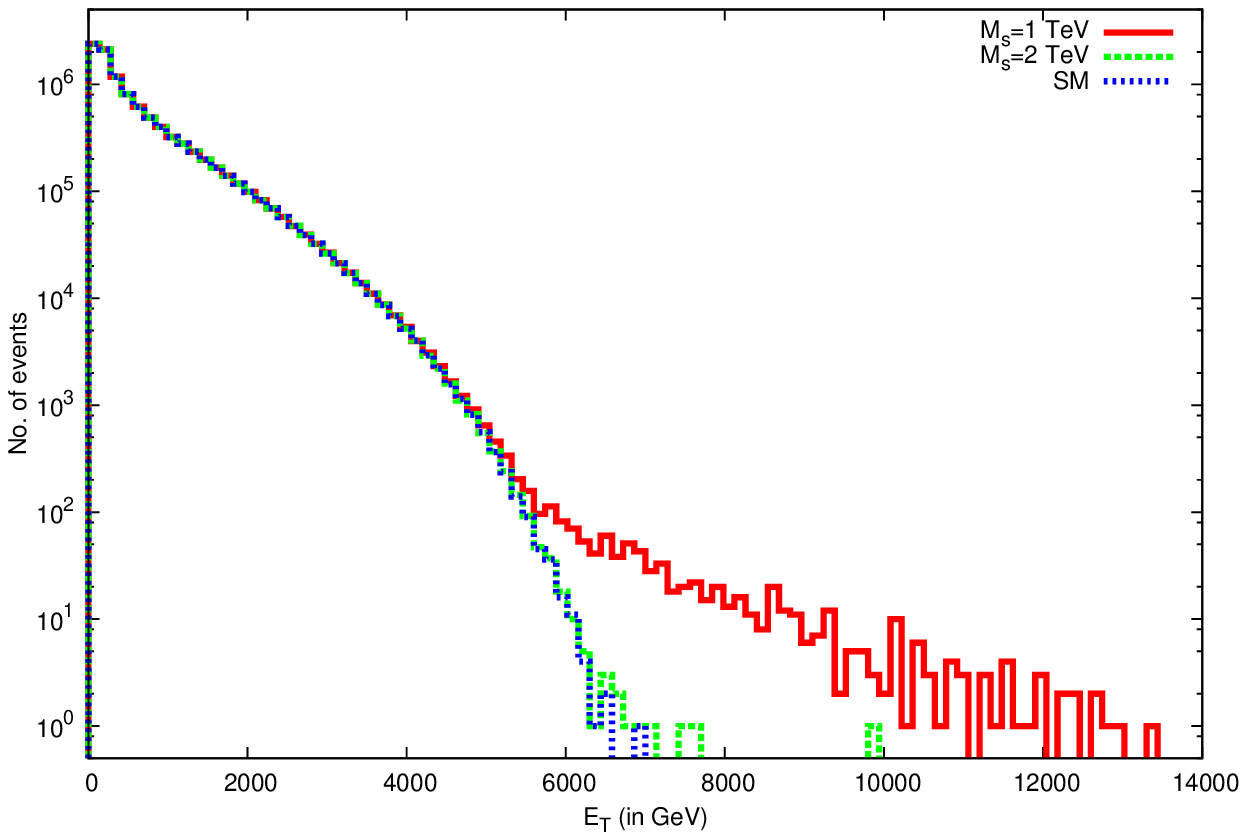} 
    \null\hfill
    \includegraphics*[width=0.5\textwidth]{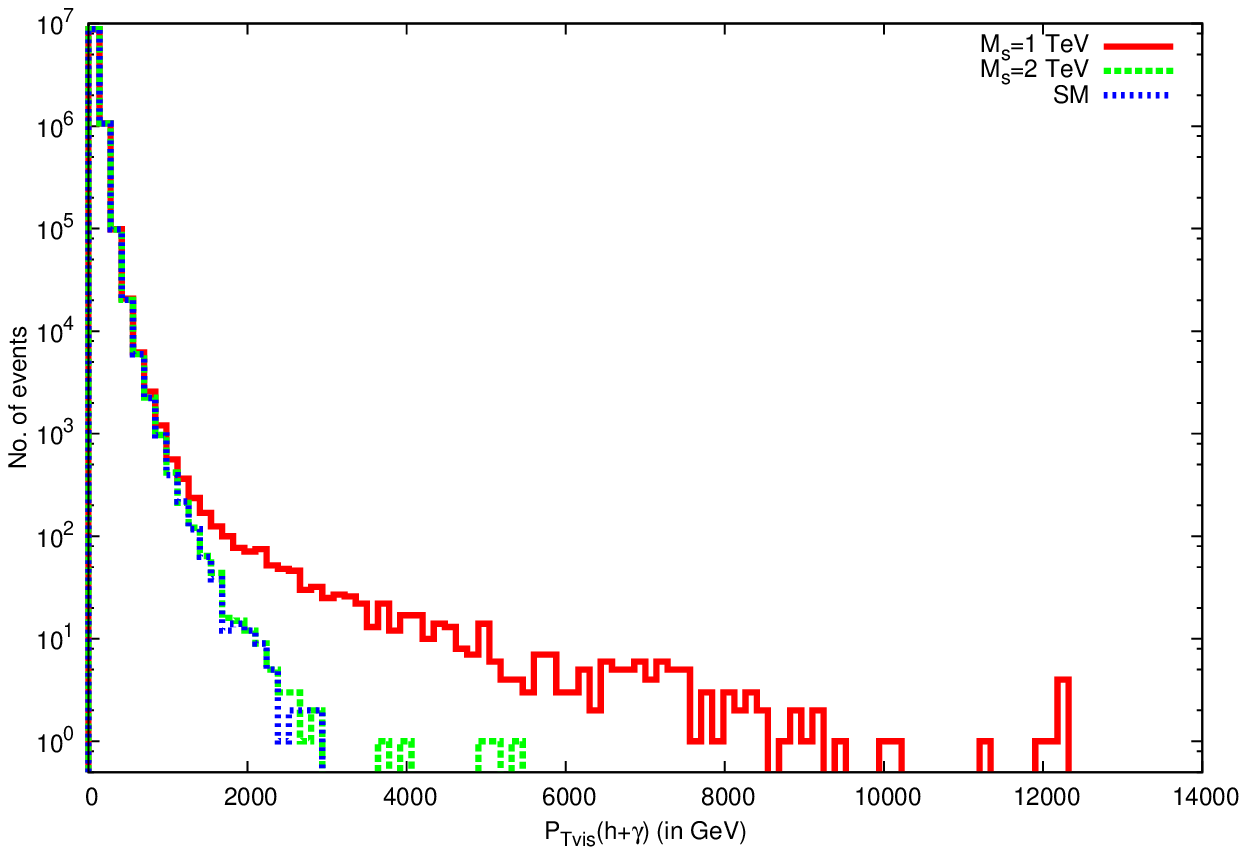} 
    \null\hfill
    }
\caption{Left Panel: Visible energy distribution for string+SM and SM-only events. The result for string
resonances is shown by the solid red histogram ($M_s$=1 TeV) and by the dashed green histogram ($M_s$=2 TeV).
String events can be identified from the high-$E_T$ tail for $M_s$=1 TeV. Right Panel: Distribution of visible
$P_T$ for $\gamma$+hadrons. The high-$P_T$ tail is a strong indicator of the presence of string resonances.}
\label{fig1}
\end{figure*}

Figure~\ref{fig1} shows the visible energy (left panel) and the transverse momentum of hadrons+photons (right
panel) for 10 million string+SM and SM-only events. The visible energy and the transverse momentum are produced by
the hard photons and the jets of the string decay. Their distributions are characterized by a long tail at high
energy/momentum. The observation of events with visible energy (transverse momentum) greater than 6~(3) TeV would
provide strong evidence of the formation of a string resonance. 
\begin{figure*}[ht]
\centerline{\null\hfill
    \includegraphics*[width=0.33\textwidth]{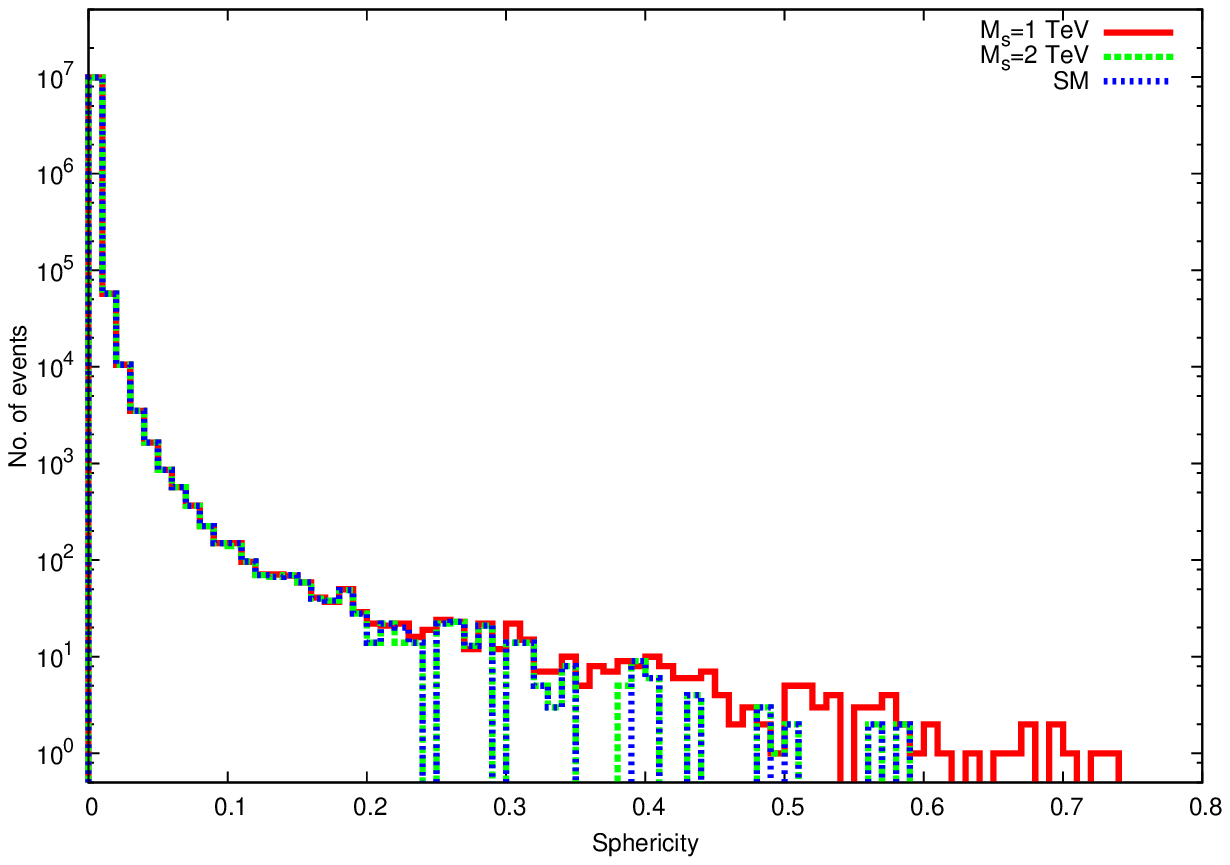}
    \null\hfill
    \includegraphics*[width=0.33\textwidth]{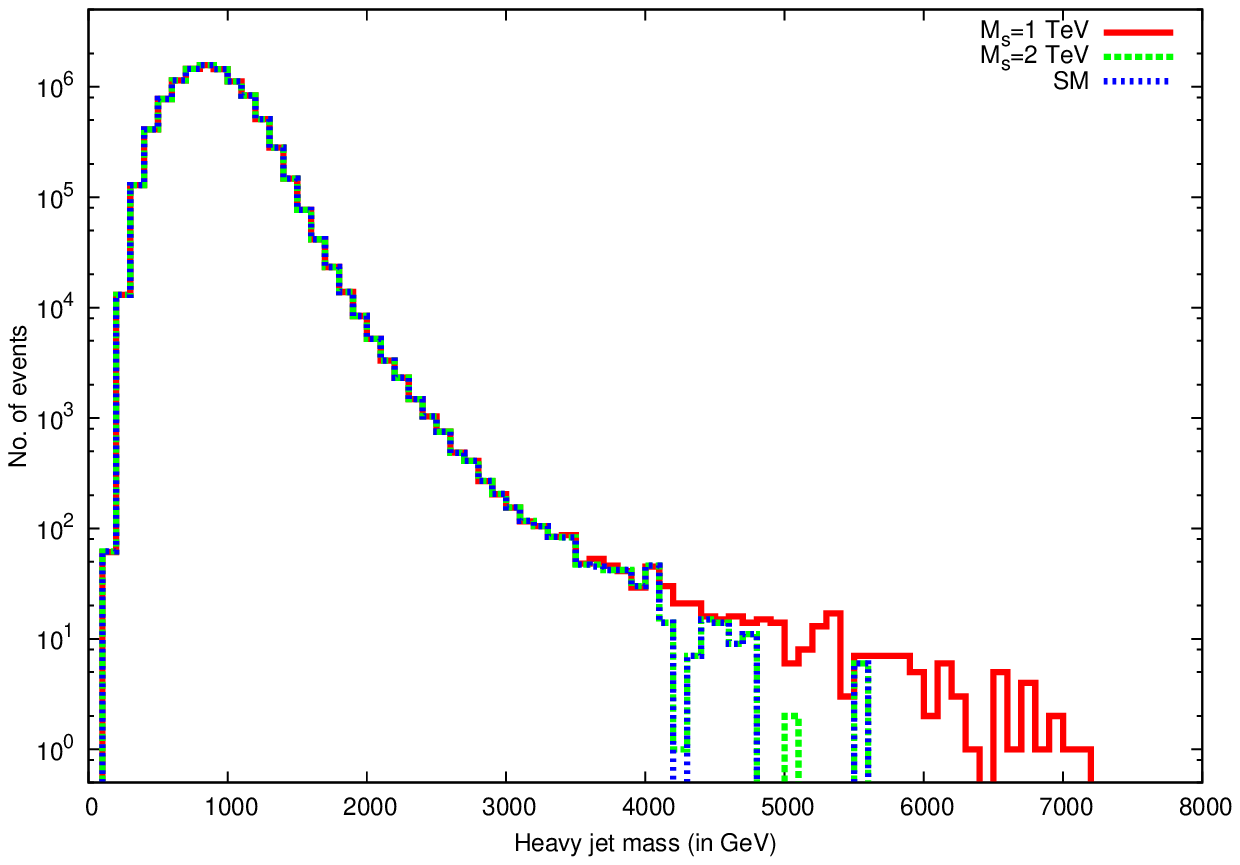}
    \null\hfill
    \includegraphics*[width=0.33\textwidth]{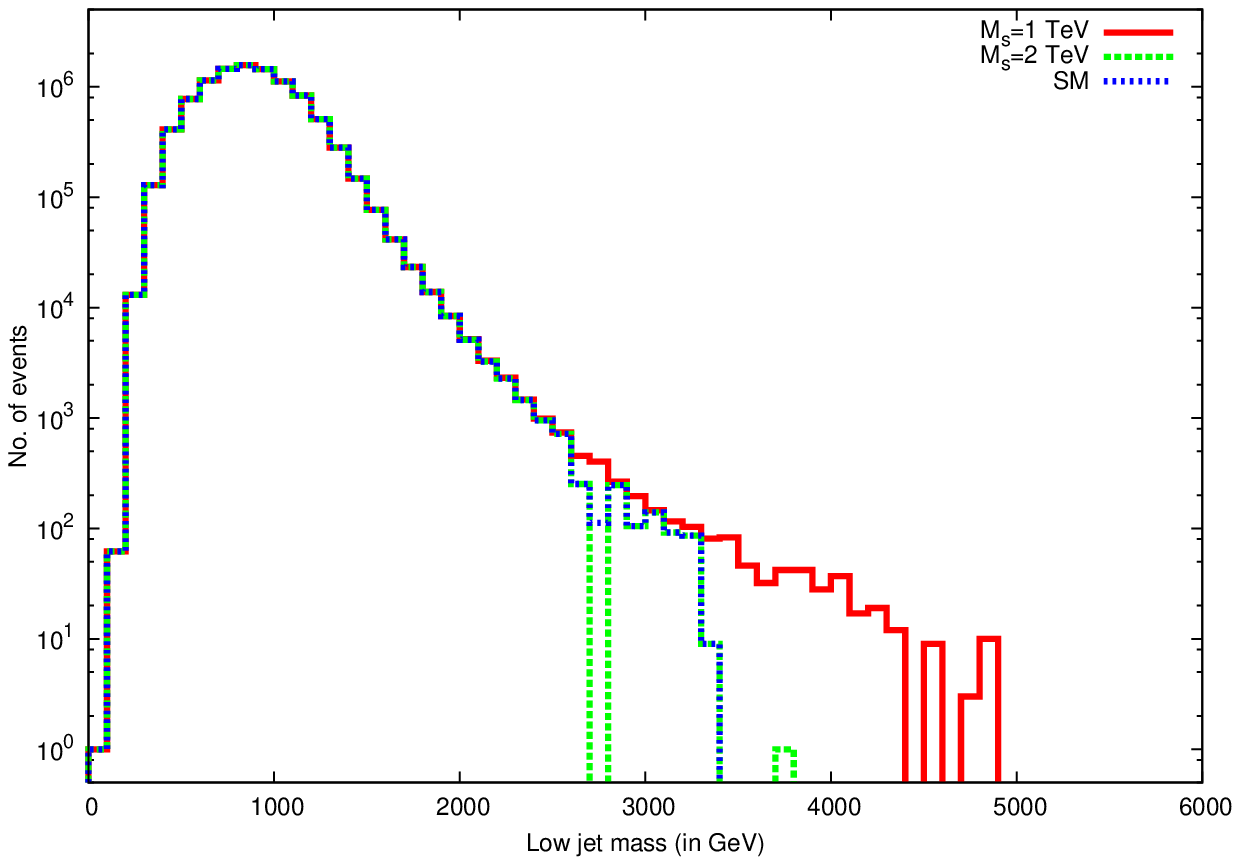}
    \null\hfill
    }
\caption{Histograms of event shape variables for 10 million string+SM and SM-only events. String events are shown
in solid red  ($M_s$=1 TeV) and dashed green ($M_s$=2 TeV). SM events are shown in dotted blue. String events have
on the average higher sphericity than SM events due to the slight increase in the number of jets (left
panel). Similar conclusions are reached from the heavy and low jet mass distributions (middle and right panel,
respectively).}
\label{fig2}
\end{figure*}

Figure~\ref{fig2} shows histograms for different event shape variables. String+SM interactions generally produce a
distribution of high $P_T$ jets at slightly higher values than the SM background, i.e. string events tend to be
more spherical than SM events. The jets originate from the decay of string resonances into photons and
hadrons. The SM generates less heavier jets than string resonances. This is evident from the middle and right
panels of Fig.~\ref{fig2}.

In the analysis of dijets, the jets are selected according to the following criteria. The detector is assumed to have
an absolute value of pseudorapidity $\eta=-\ln\{\tan(\frac{\theta}{2})\}$~=~2.6. This ensures that the jets are
originated in the hard $2\times2$ scattering rather than in multiple interactions or from the beam remnants. The
contribution of jets which do not originate in the hard scattering are minimized by fixing the the minimum transverse
energy of all particles comprising the jet ($\Sigma_i E_{T_i}$) to  40 GeV \cite{Gupta:2007cy}. The particles of the
jet must be within a cone of $R=\sqrt{(\Delta\eta^2+\Delta\phi^2)}$~=~0.5 from the jet initiator, where $\theta$ and
$\phi$ are the azimuthal and polar angles of the particle $w.r.t$ the beam axis, respectively. 
\begin{figure*}[ht]
\centerline{\null\hfill
    \includegraphics*[width=0.5\textwidth]{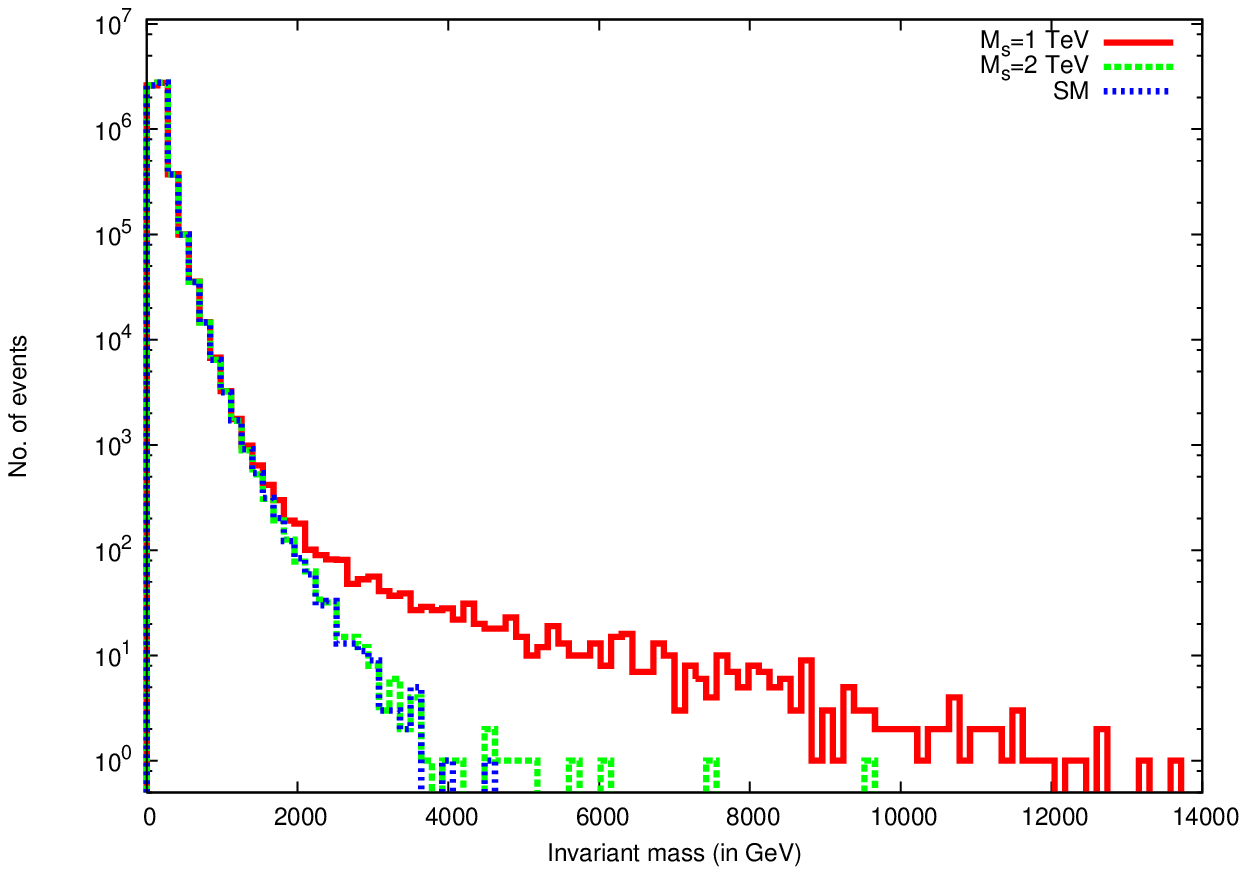}
    \null\hfill
    \includegraphics*[width=0.5\textwidth]{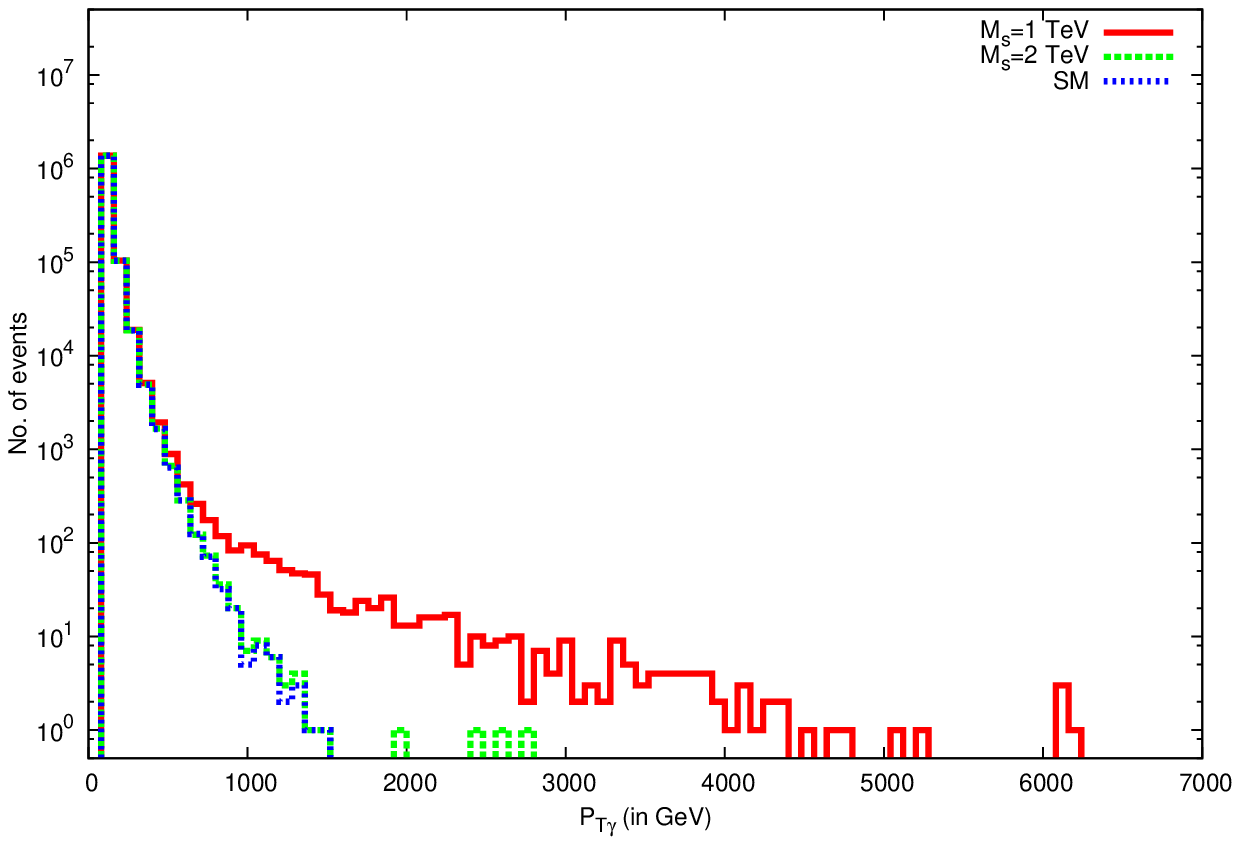}
    \null\hfill
    }
\caption{Left Panel: Dijet invariant mass distribution. String decays may result in a large invariant mass. Right
Panel: Distribution of the highest $P_{T_{\gamma}}$ for 1.5$\times10^6$ events. $\gamma$'s with high $P_T$ created in
the string decay are the source of the long tail.}
\label{fig3}
\end{figure*}

The left panel of Fig.~\ref{fig3} shows the invariant mass plot of the two jets with highest $P_T$ in each event.
Due to the nature of the interaction, the bulk of the events (both string+SM and SM) are
comprised of dijets. These were selected using the above cuts. The jet invariant mass is
\begin{equation*}
M_{12}=\sqrt{m_{1}^2+m_{2}^2+2(E_{1}E_{2}-\bar{p}_{1}.\bar{p}_{2})},
\end{equation*}
where $m_{1(2)}, E_{1(2)}~\hbox{and}~\bar{p}_{1(2)}$ are the mass, energy and momenta of jet 1~(2), respectively.
As expected, the SM invariant mass distribution is negligible beyond $\sim$ 4 TeV. This is due to the production
of direct soft photons and jets from the SM interaction. The string+SM distribution is characterized by a long
tail up to energies of several TeV (three times more than the SM). This tail is originated from the decay of
string resonances into hard jets and photons. Therefore, the measure of a large invariant mass could provide
strong evidence of a string-mediated interaction. 

The right panel of Fig.~\ref{fig3} shows the distribution of the highest $P_{T_{\gamma}}$  of isolated photons for
string+SM and SM-only events.  Following Ref.~\cite{Gupta:2007cy}, the cuts on the photon are $P_{T_{\gamma}}\ge$
80 GeV, $\eta < $ 2.6 and an isolation cut $\Sigma_n P_T <$ 7 GeV in a cone of $R$~=~0.4.  The photons from string
resonances are expected to have a higher $P_{T_{\gamma}}$ than the SM photons because they are the direct products
of the string decay. The main sources of background for direct photons are jet fluctuations  and photons
originating from the initial and final state radiation \cite{Gupta:2007cy}. In the former case, a jet consists of
a few particles including high-$P_T$ mesons (generally $\pi_0$ \cite{Gupta:2007cy}). The pions decay into a pair
of photons with a $\sim$ 99\% branching ratio. Due to the high boost, the photons have a relatively small angular
separation and therefore ``fake" a single photon in the electromagnetic calorimeter. The rate of this process is 1
out of $\sim 10^3~\hbox{to}~10^4$ events \cite{Gupta:2007cy}. Other sources of fake photons are
$H\rightarrow\gamma\gamma$ \cite{Pieri:2006bm} or processes from other exotic phenomena, e.g. SUSY
\cite{Abulencia:2007ut} or large extra dimensions \cite{ArkaniHamed:1998rs}. Isolation cuts on the photon can
effectively reduce the number of fake photons. 
\section{Conclusions\label{concl}}
We have investigated string resonances at the LHC and shown how to differentiate them from the SM background. Our
analysis has proven that string resonances could be detected when $M_s~\sim$ 1 TeV. String events show higher
sphericity and higher visible energy than the SM background. These quantities allow discrimination of string
events from SM background when combined with the measure of the $P_T$ of isolated photons and the dijet invariant
mass. Since the final products of the string resonances are directly produced from the string decay, the dijet
invariant mass is characterized by a tail at high energies which is absent in the SM.

Other exotic phenomena could also be observed at the LHC near the TeV scale, such as the formation of BHs. A
powerful way of discriminating between BH and string events would be searching for a $Z_0$ mass peak in the
invariant mass of high-$P_T$ leptons. $Z_0$ production is highly suppressed in case of string events
\cite{Anchordoqui:2008ac}. On the contrary BH decay is characterized by the production of a variety of particles
with high transverse momentum. A rough counting of the number of degrees of freedom of these particles shows that
the estimated rate of hadron-to-lepton production is 5:1 and the rate of $Z_0$ and $\gamma$ production is
comparable ($\sim$ 2\% to 3\%) with the $Z_0$ bosons decaying into opposite-sign leptons with a 3.4\% branching
ratio. Thus the invariant mass distribution of BH events peaks at $\sim$ 92 GeV, confirming the production of a $Z_0$
boson \cite{Roy:2008we}. The presence of a peak at $\sim$ 92 GeV in the invariant mass of leptons would
effectively rule out formation of string resonances.
\section*{Acknowledgments} 
The authors would like to thank L.~A.~Anchordoqui for his many valuable suggestions and comments.

\end{document}